\documentclass[preprint,nofootbib]{revtex4-1}

\newcommand{\gf}[1]{\mathbb{F}_{#1}}
\usepackage{amsmath,amssymb,mathrsfs}
\usepackage{bm}
\usepackage{makeidx}
\usepackage{txfonts}
\usepackage{color}
\usepackage[dvipdfmx]{graphicx}
\begin{document}
\begin{titlepage}
\title{The Nakano--Nishijima--Gell-Mann Formula From Discrete Galois Fields}
\author{
Keiji Nakatsugawa$^{1,2,*}$\footnote[0]{* Corresponding author}, 
Motoo Ohaga$^{1,2}$,
Toshiyuki Fujii$^{2,3}$, 
Toyoki Matsuyama$^{2,4}$ and
Satoshi Tanda$^{1,2,*}$
}
\affiliation{
$^{1}$Department of Applied Physics, Hokkaido University, Sapporo 060-8628, Japan
\\
$^{2}$Center of Education and Research for Topological Science and Technology, Hokkaido University, Sapporo 060-8628, Japan
\\
$^{3}$Department of Physics, Asahikawa Medical University, Asahikawa 078-8510, Japan
\\
$^{4}$Department of Physics, Nara University of Education, Takabatake-cho, Nara 630-8528, Japan
}
\date{\today}
\begin{abstract}
The well known Nakano--Nishijima--Gell-Mann (NNG) formula relates certain quantum numbers of elementary particles to their charge number. This equation, which phenomenologically introduces the quantum numbers $I_z$ (isospin), $S$ (strangeness), etc., is constructed using group theory with real numbers $\mathbb{R}$. But, using a discrete Galois field $\gf{p}$ instead of $\mathbb{R}$ and assuring the fundamental invariance laws such as unitarity, Lorentz invariance, and gauge invariance, we derive the NNG formula deductively from Meson (two quarks) and Baryon (three quarks) representations in a unified way. 
Moreover, we show that quark confinement ascribes to the inevitable fractionality caused by coprimeness between half-integer (1/2) of isospin and number of composite particles (e.g., three).
\end{abstract}
\nopagebreak
\maketitle
\end{titlepage}

\section{Introduction}

The standard model of particle physics has been established based on physics with real numbers ($\mathbb{R}$). However, problems such as the problem of infinite in gravity are yet unsolved. To~avoid such problems, quantum gravity theories like superstring theory or loop quantum gravity are developing, but~neither of those theories has been completed. On~the other hand, the~assumption that all of the physics in the universe is constructed solely of $\mathbb{R}$ has not been proved. For~instance, Yukawa considered extended wave function of elementary particles using ``elementary domains'' \cite{Yukawa} and Snyder~\cite{Snyder} and Finkelstein~\cite{Finkelstein} simply considered quantization of space-time. However, these approaches are incomplete and a unified formulation is unknown. Here, we assert that a different mathematical structure, namely a world with a finite~\cite{Luminet} and discrete space-time~\cite{Coish,Nambu,Jarnefelt2,Beltrametti,Shapiro,Joos,Ahmavaara,
Morris,Lev2010,Lev2020}, is needed for the following~reasons.

The description of a discrete space-time would require a discrete mathematical structure. Unitary transformation from quantum mechanics and Lorentz (M\"obius) transformation from relativity requires this structure to be a number field, where addition, subtraction, multiplication, and~division can be defined. The~set of integers $\mathbb{Z}$ is not a field because division is not defined. For~instance, fractional numbers such as $1/3$ are not defined. What we would consider in this paper is called a Galois field ($\gf{p}=\mathbb{Z}/p\mathbb{Z}$), that is, the~set of integers where two numbers differing by a multiple of a prime number $p$ are equivalent~\cite{Coish,Nambu,Jarnefelt2,Beltrametti,Shapiro,Joos,Ahmavaara,
Morris,Lev2010,Lev2020}. $\gf{p}$ has a finite, periodic, and~discrete structure. 
Mathematical and physical justification to consider this field are as~follows.

Mathematically, there is another system called p-adic numbers $\mathbb{Q}_p$ ($p:$ prime) that is equivalent to $\mathbb{R}$ \cite{Volovich}. $\mathbb{Q}_{p}$ is regarded as the limit of p-adic integers $\mathbb{Z}/p^{n}\mathbb{Z}$ with $n\to\infty$.
\begin{equation*}
\mathbb{Z}/p\mathbb{Z}, \mathbb{Z}/p^{2}\mathbb{Z}, \mathbb{Z}/p^{3}\mathbb{Z}, \cdots, \mathbb{Z}/p^{n}\mathbb{Z}\overset{n\to \infty} \longrightarrow \mathbb{Q}_p.
\end{equation*}
{Moreover, In  adele ring theory}, $\mathbb{R}$ is regarded as the limit of $\mathbb{Q}_{p}$ with $p\to\infty$ \cite{suuron}. 
\begin{equation*}
\mathbb{Q}_2, \mathbb{Q}_3,\mathbb{Q}_5,\cdots,\mathbb{Q}_{p}\overset{p\to\infty}\longrightarrow  \mathbb{R}.
\end{equation*}
{So, from~the viewpoint }of number theory, the~geometry constructed by the real numbers $\mathbb{R}$ is only an approximation of a finite and discrete geometry~\cite{Penrose,Ahmavaara}. 
Physics over $\mathbb{R}$ may be obtained from physics over $\gf{p}$ under the appropriate~limit.

Physically, Galois fields can appear naturally in our universe from multiple perspectives. 
First, recent observation of microwave background radiation suggests that the world is finite and periodic ($I_{120}$: Poincar\'e Dodecahedron) \cite{Luminet}. 
Second, if~space-time has the Planck length as a characteristic scale, then Lorentz transformation must be discrete~\cite{DSR}. Theoretically, Jarnefelt estimated a huge but finite number of world coordinates as $10^{10^{81}}$ \cite{Coish,Jarnefelt2,Jarnefelt}. 
Third, the~theory of quantum gravity needs to be expanded because of the difficulty of renormalizability. This problem would not appear in $\gf{p}$ in the first place~\cite{Coish,Nambu,Lev2010,Lev2020}, so the expansion is expected to be achieved by using a discrete system.  
In~standard particle physics theory, the~existence of antiparticles depends on additional assumptions, while in the Galois field such strange assumptions are not necessary~\cite{Lev2010}. 
Fourth, the~field theory of Galois field was investigated by Nambu in the context of recurrence time~\cite{Nambu}. Recently, the~discreteness of time is also discussed the context of time crystals~\cite{CTC,Wilczek,TC,TO,Sacha}. So, if~space-time is both finite and discrete, we have to re-examine the laws of physics~\cite{Nambu}.

In this article, we re-examine a previous model and classify hadrons by using a discrete Galois field $\gf{p}$. When we reconstruct new theories with a Galois field, these new theories must satisfy fundamental conservation laws, namely those related to unitarity, Lorentz invariance, and~gauge invariance. Consequently, instead of the Nakano--Nishijima--Gell-Mann (NNG) formula (which is proposed using continuous coordinate $\mathbb R$), we obtained the alternate formula $Q=2(n+I)$, where $Q$ is charge number, $n$ is multivaluedness in Galois field, and~$I$ is total isospin. This formula is derived from Meson (two quarks) and Baryon (three quarks) representations in a unified way. The~use of isospin was not investigated in detail by previous authors~\cite{Coish,Jarnefelt,Jarnefelt2,Beltrametti,Shapiro,Joos,Ahmavaara, Kustaanheimo,Morris} who introduced finite geometry into the field theory. Since the NNG formula also applies to individual quarks, we apply the new formula to up-quark and down-quark, which inevitably lead to fractional numbers. Fractional numbers in Galois fields are, however, very large numbers. Therefore, isolated quarks with fractional charge can have a very large energy. We surmise that \emph{quark confinement} ascribes to this fractionality in Galois field. These results may be a starting point to develop a theory without problems of infinity.

\section{Galois~Field}
Here, we introduce the Galois fields $\gf{p}$ and $\gf{p^2}$. A~number field is a mathematical structure where addition, subtraction, multiplication and division are defined. Let $\mathbb{Z}/m\mathbb{Z}$ be the set of integers where two numbers differing by an integral multiple of $m\in\mathbb{Z}$ are regarded equivalent. A~well-known result from number theory is that $\mathbb{Z}/m\mathbb{Z}$ is a number field if and only if $m$ is a prime number. This finite field is called a Galois field which is denoted by $\gf{p}$:
\begin{equation}
\gf{p}=\{0,1,2,\dots,p-1\}.
\end{equation}

A Galois field is not an ordered field. 
But, as~was made clear in Reference~\cite{Kustaanheimo}, $\gf{p}$ can be partially ordered if $p$ has the form
\begin{equation}
p=8x\prod_{i=1}^kq_i-1
\label{SpecialP}
\end{equation}
where $x$ is an odd integer and $\prod_iq_i$ is the product of the first $k$ odd primes. So, the~usual notion of magnitude (large and small) and sign (positive and negative) hold for the first $N\sim q_k\sim \mathrm{ln}p$ elements. Therefore, we consider a prime of the form in Equation \eqref{SpecialP} which is assumed to be large enough to construct a geometry containing all coordinates in the~universe.

Quantum mechanics requires complex numbers. To~consider complex numbers in a Galois field, it is natural to take an analogue of the ordinary extension of real numbers to complex numbers. That is to say, the~set of ``complex" numbers
\begin{align}
z&=x+iy,\quad (x, y \in \gf{p})
\\
i^{2}&=-1
\end{align}
is a ``complex" Galois field which is denoted by $\gf{p^{2}}$.
For a prime $p$ of the form $p=4n$ $-$ 1 ($n\in\mathbb{Z}$), complex conjugation is given by $i^{p} = -i = i^\ast$ and $z^{p} = x - iy = z^\ast$ since we have
\begin{equation}
(x+iy)^p=x^p-iy^p=x-iy.
\end{equation}
The first equality holds because $p$ is a character of the field, that is $p\cdot a=0$ for all $a\in\gf{p}$; The~second equality follows from Fermat's little theorem $a^{p}\equiv a\mbox{ mod }p$. Note that $|z|^2$ can become negative for the non-ordered part of $\gf{p}$.

\section{Lorentz Group in a Galois Field: The Coish~Group}
Lorentz transformation can be recovered with the Galois field introduced in the previous section~\mbox{\cite{Coish,Ahmavaara,Beltrametti,Morris}}. Coordinates in Minkowski space-time are closely related to spinor structure through the Pauli matrix representation
\begin{equation}
X=\left(
\begin{matrix}
x_0+x_3 & x_1-ix_2 \\
x_1+ix_2 & x_0-x_3
\end{matrix}
\right)=x_0{\bf 1}+x_1\sigma_1+x_2\sigma_2+x_3\sigma_3
\end{equation}
where $x_0,x_1,x_2,x_3\in\gf{p}$ and
\begin{align*}
\sigma_1&=\left(
\begin{matrix}
0 & 1 \\
1 & 0
\end{matrix}
\right),
\quad
\sigma_2=\left(
\begin{matrix}
0 & -i \\
i & 0
\end{matrix}
\right),
\quad
\sigma_3=\left(
\begin{matrix}
1 & 0 \\
0 & -1
\end{matrix}
\right)
\end{align*}
are Pauli matrices. (See Reference~\cite{Weinberg} for the case of $x_\mu\in\mathbb{R}$.) The metric form is given as the determinant of $X$:
\begin{equation}
\|X\|=x_0^{2}-x_1^{2}-x_2^{2}-x_3^{2}.
\label{QuadraticForm}
\end{equation}

Let $a$ be a 2 $\times$ 2 matrix with coefficients in $\gf{p^2}$. The~linear transformation
\begin{equation}
X^{'}=a^{\dagger}Xa
\label{LorentzTransformation}
\end{equation}
is a proper Lorentz transformation if it leaves the metric form Equation \eqref{QuadraticForm} invariant, that is, if~$\|a\|^\ast\|a\|=\|a\|^{p+1}=1$ holds. Such matrices form the orthogonal group SL$(2,\gf{p^2})$. In~finite geometry, however, there also exist transformations with $\|a\|^{p+1}=-1$ which reverse the sign of the metric~\cite{Coish}.

\begin{figure}[h]
\begin{center}
\includegraphics[width=0.4 \linewidth]{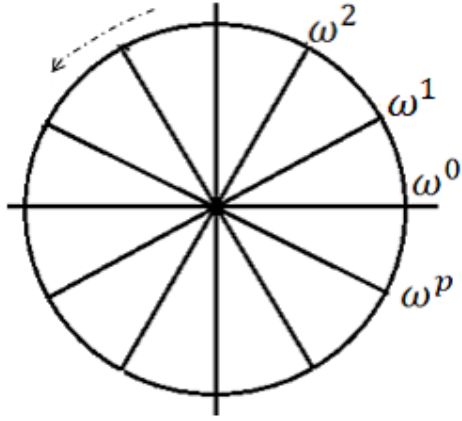}
\end{center}
\caption{Schematic representation of the discrete multivaluedness with phase factor $\omega$. Each set of $\gf{p^2}$ constitutes a  cyclotomic field. When $p=4n-1$, the~field has $p+1$ elements. A~Lorentz group in a Galois field has $p+1$ multivaluedness such as $\omega^{\alpha}a,\ \alpha=0,1,\cdots,p$. Regarding $\omega$ as a phase factor, it constitute a global gauge transformation in a Galois~field.}
\label{omega}
\end{figure}

In addition, suppose that matrices $a_1$ and $a_2$ lead to the same Lorentz transformation. Then~\mbox{$U\equiv a_{1}a_{2}^{-1}$} transforms $X$ as
\begin{equation}
U^{\dagger}XU=X.
\label{UnitaryTransformation}
\end{equation}
Since the set of $X$ contains the identity matrix, Equation \eqref{UnitaryTransformation} leads to $U^\dagger U=UU^\dagger=1$. So, $U$ is a~unitary matrix of the form
\begin{equation}
U=\lambda \bf{1}.
\end{equation}
Given that $U$ is unitary, $\lambda$ satisfies \\
\begin{equation}
\lambda^{*}\lambda=1.
\end{equation} 
In a Galois field, such $\lambda$ become discrete and multivalued. That is, for~a prime $p$ of the form $p=4n-1$ (such as in Equation \eqref{SpecialP}) we have
\begin{equation}
\lambda=\omega^{\alpha}\ ,\ a_{1}=\omega^{\alpha}a_{2}.
\end{equation}
where $\alpha=0,\ 1,\ \cdots,\ p$. So, the~the orthogonal group in a Galois field has $p+1$ multivaluedness as with a cyclotomic field, and~$\omega$ acts as a phase factor (Figure \ref{omega}). In~this way, we obtain the full Coish group. The~spinor group, which is a subgroup of the rotational group, has the same multivaluedness. This multivaluedness will plays an essential role for the classification of~hadrons.

\unskip

\section{Gauge Transformation in Galois~Field}
In this section, the~charge number $Q$ is derived with gauge transformation in Galois field in accordance with the derivation given in Reference~\cite{Coish}. Because~of homomorphism, an~orthogonal group leads a spinor group. The~irreducible representation of the orthogonal group must be a matrix with elements of $\gf{p^2}$ as its coefficients. The~spinor group is a subgroup of of the two dimensional general linear homogeneous Galois group $GLH(2,p^{2})$ over the complex finite field $\gf{p^2}$. The~irreducible representations of $GLH(2,p^{2})$ are given by Brauer and Nesbitt~\cite{Brauer_Nesbitt} and extended by Coish~\cite{Coish} as
\begin{equation}
\|a\|^na^{(2j)}\otimes {\bar a}^{(2k)},
\label{twin-direct-product}
\end{equation}
where
\begin{align}
j,k&=0,\ 1/2,\ 1,\ 3/2, \cdots (p-1)/2,
\\
n&=0,\ \pm1/2,\ \pm1, \cdots \pm(p+1/2),\ p+1.
\label{nvalues}
\end{align}
Here, $a$ is a spinor matrix which transforms a spinor
\begin{equation}
\left(\begin{matrix}\psi_1\\ \psi_2\end{matrix}\right)
\to
\left(\begin{matrix}\psi'_1\\ \psi'_2\end{matrix}\right)
=
a
\left(\begin{matrix}\psi_1\\ \psi_2\end{matrix}\right).
\end{equation}
$\|a\|^n$ is interpreted as the multivaluedness of the representation. Next, consider the polynomial
\begin{equation}
f(\psi)=\sum_{m=-j}^jf_m\psi_1^{j+m}\psi_2^{j-m}.
\end{equation}
The set $\left\{\psi_1^{j+m}\psi_2^{j-m}\right\}$ can be regarded as the linearly independent set of basis for a \mbox{$(2j+1)$-dimensional} vector space. The~action of $a$ on $f(\psi)$ is given by
\begin{equation}
f(a\psi)=\sum_{m=-j}^jf_m(\psi_1')^{j+m}(\psi_2')^{j-m}=\sum_{m=-j}^jf_ma^{(2j)}_{mn}\psi_1^{j+n}\psi_2^{j-n}.
\end{equation}
So, $a^{(2j)}$ is a matrix transforming the monomials $\psi_1^{j+m}\psi_2^{j-m}$. In~particular, the~phase factor $\omega^\alpha$ transforms $a\to\omega^\alpha a$ and $a^{(2j)}\to \omega^{2j\alpha}a^{(2j)}$. This phase factor corresponds to a gauge transformation, and~Equation \eqref{twin-direct-product} is transformed as follows
\begin{equation}
\|a\|^na^{(2j)}\otimes \bar{a}^{(2k)}\longrightarrow\omega^{Q\alpha}\|a\|^na^{(2j)}\otimes \bar{a}^{(2k)},
\end{equation}
where
\begin{equation}
Q=2(n+j-k).
\label{Q=2(n+j-k)}
\end{equation}

Therefore, we naturally derive Equation \eqref{Q=2(n+j-k)} as the charge number reflecting  discreteness of gauge transformation in Galois~field.

\section{Classification of~Hadrons}
Now, we can classify Hadrons in a Galois field. Coish interpreted $j$ and $k$ as spins and represented baryons and mesons with Equations \eqref{twin-direct-product} and \eqref{Q=2(n+j-k)}. However, his model does not agree with the standard model of particle physics. So, we give a different interpretation. 
Instead of spins, we interpret $j$ and $k$ as isospin. More precisely, to~encompass the multiplet structure of hadrons, it would be appropriate to interpret $j,k$ as the total isospin $I$ instead of $I_z$. Mesons, which are composed of particles and anti-particles, are transformed by the representation
\begin{equation}
D^{(j,-k,n)}_{\text{meason}}\equiv\|a\|^na^{(2j)}\otimes {\bar a}^{(2k)},
\label{MesonRep}
\end{equation}
but it is now interpreted as a representation of $GLH(2,p^{2})\times GLH(2,p^{2})^\ast$. Baryons, which are composed of three quarks, are transformed by another representation
\begin{equation}
D^{(i,j,k,n)}_{\text{baryon}}\equiv\|a\|^na^{(2i)}\otimes a^{(2j)}\otimes a^{(2k)}
\label{BaryonRep}
\end{equation}
of $GLH(2,p^{2})\times GLH(2,p^{2})\times GLH(2,p^{2})$. Here, the~contribution of strangeness appears as a modification of the multivaluedness $n$.

Mesons and baryons can be classified according to the charge number
\begin{equation}
Q=2(n+I),
\label{Coishmodel}
\end{equation}
where $I=j-k$ for mesons and $I=i+j+k$ for~baryons.

Tables~\ref{meson table} and \ref{baryon table}, respectively, show quantum numbers and representations of mesons and baryons in a Galois field. Here, we did not use $\mathrm{SU}(3)$ symmetry but assume the known quark~contents.

\begin{table}[h]
\centering
\caption{Quantum numbers of mesons in a Galois field. $Q$ is the charge number of baryon and $j,k$ are total isospins of quarks. $n$ is derived from Equation \eqref{Coishmodel}.}
\label{meson table}
  \begin{tabular}{|c||c|c|c|c|c|c|} \hline
    Meson & $Q$ & $n$ & $I$ & $j$ & $k$ & Representation \\ \hline\hline
    $K^{+}$ & 1 & 0 & 1/2 & 1/2 & 0 & $\|a\|^{0}a^{(1)}\otimes \bar{a}^{(0)}$ \\ \hline
    $K^{0}$ & 0 & $-$1/2 & 1/2 & 1/2 & 0 & $\|a\|^{-\frac{1}{2}}a^{(1)}\otimes \bar{a}^{(0)}$ \\ \hline
    $\pi^{+}$ & 1 & 1/2 & 0 & 1/2 & 1/2 & $\|a\|^{\frac{1}{2}}a^{(1)}\otimes \bar{a}^{(1)}$ \\ \hline
    $\pi^{0}$ & 0 & 0 & 0 & 1/2 & 1/2 & $\|a\|^{0}a^{(1)}\otimes \bar{a}^{(1)}$ \\ \hline
    $\pi^{-}$ & $-$1 & $-$1/2 & 0 & 1/2 & 1/2 & $\|a\|^{-\frac{1}{2}}a^{(1)}\otimes \bar{a}^{(1)}$ \\ \hline
    $\bar{K^{0}}$ & 0 & 1/2 & $-$1/2 & 0 & 1/2 & $\|a\|^{\frac{1}{2}}a^{(0)}\otimes\bar{a}^{(1)}$ \\ \hline
    $K^{-}$ & $-$1 & 0 & $-$1/2 & 0 & 1/2 & $\|a\|^{0}a^{(0)}\otimes \bar{a}^{(1)}$ \\ \hline
         \end{tabular}
         \end{table}
         
\begin{table}[h]
\centering
\caption{Quantum numbers of baryons in a Galois field. $Q$ is the charge number of baryon and $i,j,k$ are total isospins of quark. $n$ is derived from Equation \eqref{Coishmodel}. } 
\label{baryon table}
  \begin{tabular}{|c||c|c|c|c|c|c|c|} \hline
     Baryon &  $Q$ &  $n$ &  $I$ &  $i$ &  $j$ &  $k$ & Representation \\ \hline\hline
    $p$ & 1 & $-$1 & 3/2 & 1/2& 1/2 & 1/2 & $\|a\|^{-1}a^{(1)}\otimes a^{(1)}\otimes a^{(1)}$\\ \hline
    $n$ & 0 & $-$3/2 & 3/2 & 1/2 & 1/2&1/2 & $\|a\|^{-\frac{3}{2}}a^{(1)}\otimes a^{(1)}\otimes a^{(1)}$\\ \hline
    $\Sigma^{+}$ & 1 & $-$1/2 & 1 & 1/2 &1/2& 0 & $\|a\|^{-\frac{1}{2}}a^{(1)}\otimes a^{(1)}\otimes a^{(0)}$ \\ \hline
    $\Sigma^{0}$ & 0 & $-$1 & 1 & 1/2 & 1/2 &0& $\|a\|^{-1}a^{(1)}\otimes a^{(1)}\otimes a^{(0)}$\\ \hline
    $\Sigma^{-}$ & $-$1 & $-$3/2 & 1 &  1/2&1/2 &0 & $\|a\|^{-\frac{3}{2}}a^{(1)}\otimes a^{(1)}\otimes a^{(0)}$ \\ \hline
    $\Xi^{0}$ & 0 & $-$1/2 & 1/2 & 1/2 & 0 &0&$\|a\|^{-\frac{1}{2}}a^{(1)}\otimes a^{(0)}\otimes a^{(0)}$ \\ \hline
    $\Xi^{-}$ & $-$1 & $-$1 & 1/2 &  1/2  &0 &0 & $\|a\|^{-1}a^{(1)}\otimes a^{(0)}\otimes a^{(0)}$ \\ \hline
         \end{tabular}
\end{table}

\begin{figure}[h]
\begin{center}
\includegraphics[width=0.9 \linewidth]{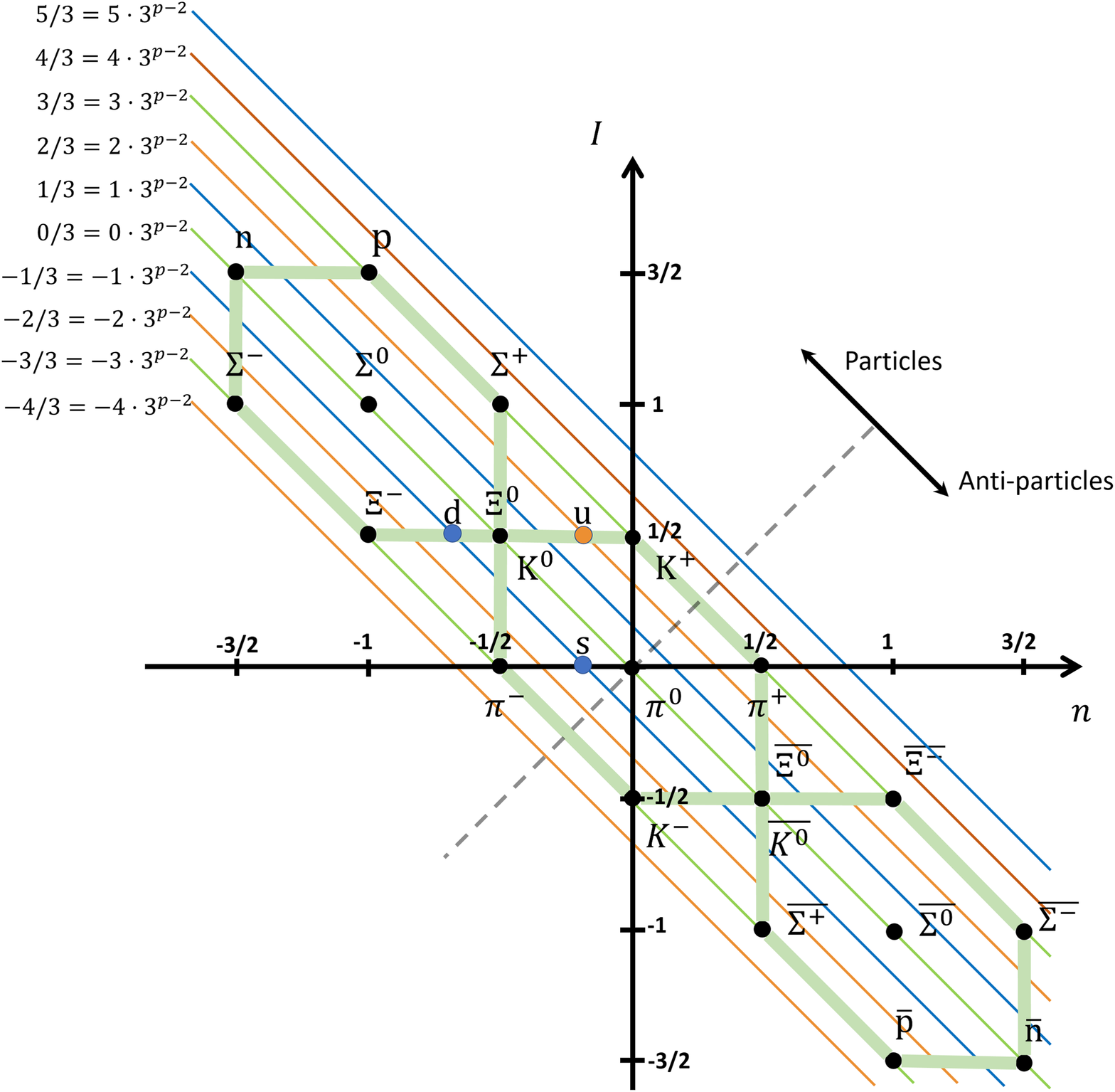}
\end{center}
\caption{Classification of hadrons with Equation \eqref{Coishmodel}. Our model take the $n$ and $I$ axis instead of $Y$ and $I_z$. We did not use $\mathrm{SU}(3)$ symmetry but assume the same quark constituent as in the standard model of particle physics. Charge numbers such as $Q=2/3=2\cdot 3^{p-2}\mbox{ mod }p$ are shown in  Figure \ref{3inv}. Note that this figure superimposes baryons, anti-bayons, and~mesons, which differ by baryon numbers. So, K\textsuperscript{0} and $\Xi$\textsuperscript{0}, for~instance, are~distinguished.}
\label{baryonmesoncomparisontable}
\end{figure}

We derived  {Equation} \eqref{Coishmodel} equivalent to the NNG formula 
\begin{equation}
Q=I_{z}+\frac{Y}{2},\label{NNG}
\end{equation}
where $Y$ is the hypercharge. Figure \ref{baryonmesoncomparisontable} shows the classification of hadrons with $n$ and $I$. Particles and anti-particles are related by CP conjugation (that is, point symmetry around the origin $n=I=0$) as in {Reference}~\cite{Coish}. 
Different values of $I$ indicate the multiplet structure of hadrons. By~comparison with the NNG formula, $n$ includes both the isospin $I_z$ and the hypercharge $Y$. In~this way, Galois fields can be used to organize the quantum number of hadrons. Similarly, as~it is well known, the~quantum number of leptons are related with an equation similar to the NNG formula; the origins of the weak isospin and weak hypercharge, which are properties of leptons introduced phenomenologically, may also be explained by the Galois~field.

More generally, we can give a unified representation for arbitrary composit particles. First, we can give a representation of a single particle as $\|a\|^na^{(2j)}$. Likewise, a~representation of particles with $N$ ($=2,3,4,5$ or higher) constituents is possible with
\begin{equation}
\bigotimes_{l=1}^N\|a\|^{n_l}a^{(2j_l)\theta_{\nu_l}}\label{generalrepresentation}
\end{equation}
where $j_l$ are half-integers, $\nu_l=0$ or $1$, $\theta_0$ is the identity map $a^{(2j)\theta_0}=a^{(2j)}$, and~$\theta_1$ maps $a^{(2j)}$ to the conjugate group $a^{(2j)\theta_1}=\bar{a}^{(2j)}$ with coefficients $\alpha^{p^1}=a^\ast$ (c.f. References~\cite{Brauer_Nesbitt,Coish}). Then, the~charge number is given by $Q=\sum_{l=1}^N2(n_l\pm j_l)$ where plus and minus correspond to $\nu_l=0$ and $\nu_l=1$, respectively. This includes a representation of mesons, baryons, pentaquark, and~so on, in~a unified~way.

\section{Quark~Confinement}
Finally, we provide a possible approach toward quark confinement which ascribes to fractionality in Galois~field.

Suppose for the moment that up-quark and down-quark can be described by the following irreducible representation of $GLH(2,p^2)$:
\begin{equation}
a_u=\|a\|^{n_u}a^{(1)},\qquad a_d=\|a\|^{n_d}a^{(1)}.
\label{udRep}
\end{equation}
We expect that proton and neutron can be expressed as $a_p=a_u\otimes a_u\otimes a_d$ and $a_n=a_u\otimes a_d\otimes a_d$, respectively. Then, the~quantum numbers of proton and neutron can be written in terms of their constituent quarks as follows:
\begin{align}
n_p&=2n_u+n_d,
\\
n_n&=n_u+2n_d,
\\
I_p&=I_n=I.
\end{align}
Substitute this decomposition into Equation \eqref{Coishmodel} and we obtain
\begin{align}
Q_p=2(2n_u+n_d+I),\quad 
Q_n=2(n_u+2n_d+I).
\end{align}
Then, we can solve for $n_u$ and $n_d$
\begin{align}
n_u=\frac{2Q_p-Q_n-2I}{6},\quad 
n_d=\frac{2Q_n-Q_p-2I}{6}.
\end{align}
Taking the values $Q_p=1$, $Q_n=0$ and $I=3/2$ as in Table \ref{baryon table} we obtain
\begin{equation}
n_u=-\frac{1}{6},\quad 
n_d=-\frac{2}{3},
\end{equation}
which are not half-integers as in \eqref{nvalues}.
Therefore, irreducible representations of quarks as in Equation~\eqref{udRep} is not~possible.

\begin{figure}[h]
\centering
\includegraphics[width=0.75\linewidth]{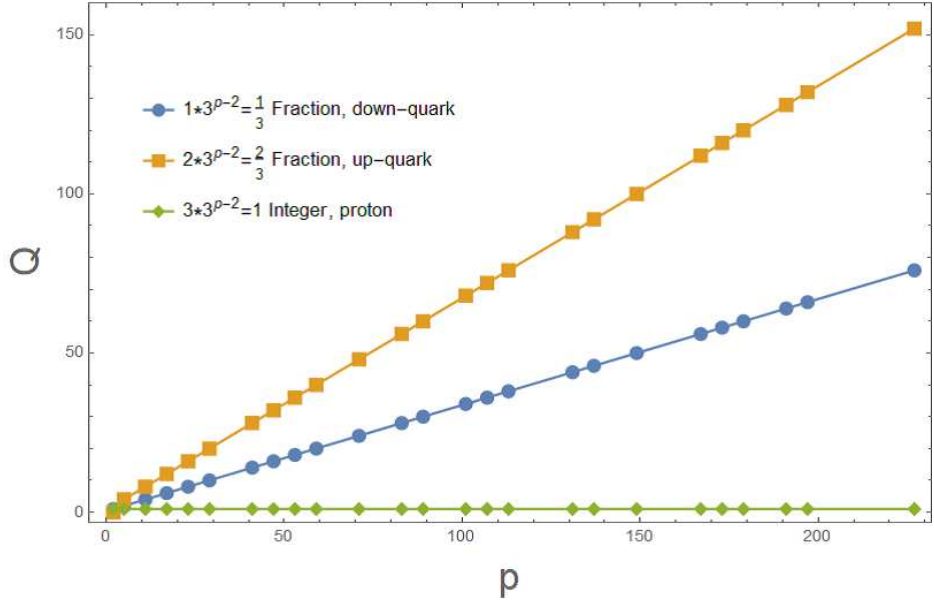}
\caption{The values of $1/3=3^{p-2}\text{ mod }p=(p+1)/3$ and $2/3=2\cdot 3^{p-2}\text{ mod }p=2(p+1)/3$ are shown for the primes of the form $p=3n-1$. Fractional numbers in a Galois field are large numbers proportional to $p$.}
\label{3inv}
\end{figure}

\begin{figure}[h]
\centering
\includegraphics[width=0.75\linewidth]{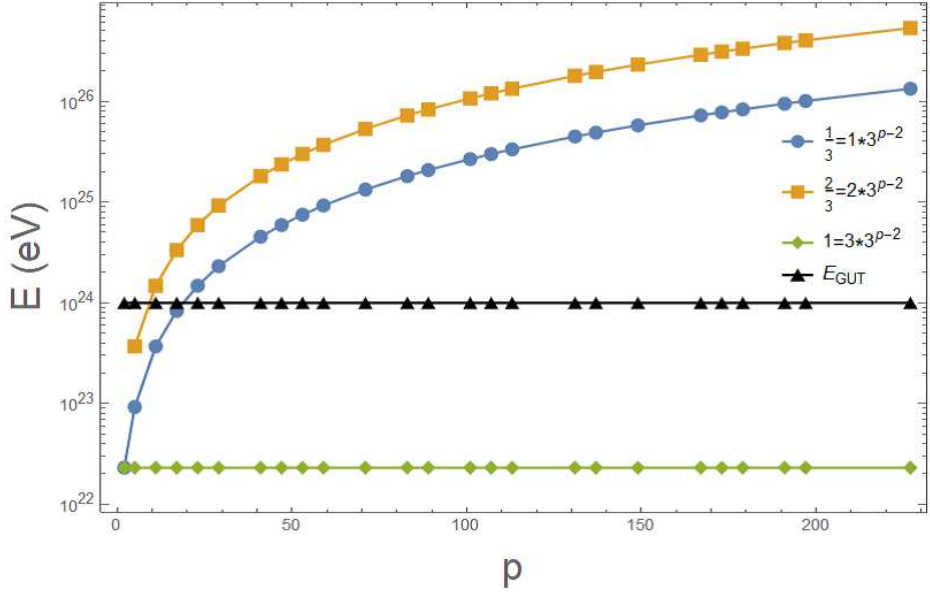}
\caption{Electromagnetic self-energy $E_\mathrm{em}=k_e\frac{Q^2}{2a}c^2$ where $Q$ is the charge number in Galois field, $k_e$ is the Coulomb constant, $a$ is the classical electron radius, and~$c$ is the speed of light. The~energy is independent of $p$ for integral charge but increase for fractional charge. The~GUT energy is shown as a rough~standard.}
\label{Energy}
\end{figure}

Now, suppose that we can nevertheless write the corresponding charge numbers as elements of $\gf{p}$. Then, we obtain $Q_u=2/3$ and $Q_d=-1/3$. Counterintuitively, these fractional quantum numbers in $\gf{p}$ are very large, which can be shown as follows. The~modular inverses in of an integer $a$ in $\gf{p}$ is defined by
\begin{equation}
a^{-1}a\equiv 1\text{ mod } p.
\end{equation}
On the other hand, from~Fermat's little theorem
\begin{equation}
a^{p-1}\equiv 1\text{ mod } p
\end{equation}
we obtain \begin{equation}
a^{-1}\equiv a^{p-2} \text{ mod } p.
\end{equation}
The prime given by Equation \eqref{SpecialP} contains the first odd prime $q_1=3$, so it has the form $p=3n-1$ for some $n\in\mathbb{N}$. In~this case, the~fractional number $1/3$ in $\gf{p}$ takes the value $3^{-1}=(p+1)/3$ (\mbox{see Figure~\ref{3inv}}). Therefore, fractional numbers in Galois field are large numbers of order $p$.
Consequently, isolated quark can have, for~instance a \textit{huge} electromagnetic self-energy (Figure \ref{Energy}). In~this case, quark confinement in $\gf{p}$ may be a consequence of energy~minimization.

\section{Discussion and~Conclusions}
Finally, we summarize and discuss our results. The~NNG formula is a phenomenological equation. 
By using a Galois field, the~NNG Formula \eqref{NNG} is rewritten as \eqref{Coishmodel}. The~new quantum number $n$, which includes isospin $I_z$ and hypercharge $Y$, arises from the multivaluedness of a linear representation of the Galois field. We also discussed quark confinement which ascribes to fractionality of Galois field. After~all, these results may indicate finiteness and discreteness of the world. While such a method is unconventional, it is not inherently inconsistent under one assumption, namely that the geometry constructed by the real numbers $\mathbb{R}$ is only an approximation of a finite and discrete geometry~\cite{Penrose,Ahmavaara}. 

As future developments, we expect that composite particles such as pentaquarks~\cite{Nakano,Pentaquark,Kim} can be explained by generalizing Equation \eqref{generalrepresentation} to any number of particles and combining the quantum numbers $n$ and $I$ to a single ``grand spin" as discussed in~\cite{Kim}. 
We expect that the inclusion of other flavors such as c (charm), t (top) and b (bottom) would require extension of our result to $GLH(3,p^2)$ (analogous to SU(3)), $GLH(4,p^2)$, and~so on, which goes in parallel to the extension of SU(2) to SU(3) or higher. Representation of particles in Figure~\ref{baryonmesoncomparisontable} can be modified accordingly. This extention can be done based on previous studies~\cite{Mark} and using a method equivalent to the ``enlarging lemma''~\cite{Kempe}.
Recently, with~the development of quantum information, Galois fields have begun to be used to construct space-time models as error-correcting codes~\cite{Holweck}. Further application of Galois fields is expected in the foundation of quantum physics~\cite{Chang} and time crystal with discrete space-time symmetry~\cite{CTC,Wilczek,TC,TO,Sacha}.

\vspace{6pt}

\acknowledgments{
We thank Kohkichi Konno, Masanori Yamanaka, Thomas Zeugmann, Tatsuhiko N. Ikeda, Masanori Morishita and Ryotaro Okazaki for stimulating and valuable~discussions.
}

\bibliography{References.bib}
\end{document}